\documentclass[3p,review]{elsarticle}

\usepackage{lineno,hyperref}
\usepackage{graphicx,amssymb,arydshln,subeqnarray,subfigure}
\usepackage{url}
\usepackage{amsbsy,amsmath,amsfonts,cases,bm,caption,color}
\usepackage[ruled]{algorithm2e} 
\modulolinenumbers[5]

\journal{Journal of the Franklin Institute}

\begin{document}

\begin{frontmatter}

\title{A semidefinite programming approach for robust elliptic localization}

\author[myprimaryaddress,mysecondaryaddress]{Wenxin~Xiong\corref{mycorrespondingauthor}}
\cortext[mycorrespondingauthor]{Corresponding author.}
\ead{w.x.xiong@outlook.com}

\author[mytertiaryaddress]{Yuming~Chen}
\ead{cymcumt@163.com}

\author[mysecondaryaddress]{Jiajun~He}
\ead{jiajunhe5-c@my.cityu.edu.hk}

\author[myquartusaddress]{Zhang-Lei~Shi}
\ead{zlshi@upc.edu.cn}

\author[mysecondaryaddress]{Keyuan~Hu}
\ead{keyuanhu2-c@my.cityu.edu.hk}

\author[mysecondaryaddress]{Hing~Cheung~So\fnref{fn3}}
\ead{hcso@ee.cityu.edu.hk}

\author[mysecondaryaddress]{Chi-Sing~Leung}
\ead{eeleungc@cityu.edu.hk}

\fntext[fn3]{EURASIP Member.}

\address[myprimaryaddress]{Department of Computer Science, University of Freiburg, Freiburg 79110, Germany}

\address[mysecondaryaddress]{Department of Electrical Engineering, City University of Hong Kong, Hong Kong, China}

\address[mytertiaryaddress]{Nanjing University of Posts and Telecommunications, College of Automation, Nanjing 210023, China}

\address[myquartusaddress]{College of Science, China University of Petroleum (East China), Qingdao 266580, China}

\begin{abstract}
	This short communication addresses the problem of elliptic localization with outlier measurements. Outliers are prevalent in various location-enabled applications, and can significantly compromise the positioning performance if not adequately handled. Instead of following the common trend of using $M$-estimation or adjusting the conventional least squares formulation by integrating extra error variables, we take a different path. Specifically, we explore the worst-case robust approximation criterion to bolster resistance of the elliptic location estimator against outliers. From a geometric standpoint, our method boils down to pinpointing the Chebyshev center of a feasible set, which is defined by the available bistatic ranges with bounded measurement errors. For a practical approach to the associated min-max problem, we convert it into the convex optimization framework of semidefinite programming (SDP). Numerical simulations confirm that our SDP-based technique can outperform a number of existing elliptic localization schemes in terms of positioning accuracy in Gaussian mixture noise.
\end{abstract}


\end{frontmatter}


\section{Introduction}
\label{Sect_Intro}

Over the last decade, elliptic localization techniques have been widely adopted in various location-based applications, driven by the contemporary rise of multistatic systems such as distributed MIMO radar, sonar, and wireless sensor networks \cite{LRui1}. Multistatic systems are typically composed of multiple spatially separated transmitters and receivers, whose positions, $\bm{t}_{m} = [x_{m}^t, y_{m}^t]^T$ (for $m=1,...,M$) and $\bm{s}_{l} = [x_{l}, y_{l}]^T$ (for $l = 1,...,L$), respectively, are known in advance. Mathematically, elliptic localization involves an estimation problem, aimed at ascertaining the location of a signal-reflecting/relaying target, $\bm{x} = [x, y]^T$, by utilizing the sensor-collected bistatic ranges (BRs):
\begin{align}{\label{meas_BR}}
	&\hat{r}_{m,l} = {\| \bm{x} - \bm{t}_{m} \|}_{2} + {\| \bm{x} - \bm{s}_{l} \|}_{2} + e_{m,l},\nonumber\\
	&~m = 1,...,M,~l = 1,...,L,
\end{align}
where ${\|\cdot\|}_{2}$ denotes the $\ell_{2}$-norm and $\{ e_{m,l} \}$ represent the BR measurement errors. Simply put, the BR is the combined distance from a transmit antenna to the target and from the target to a receive antenna. It is acquired by multiplying the complete signal propagation time from the transmitter through the target to the receiver, also known as the time-sum-of-arrival (TSOA) \cite{HCSo}, by the speed of signal propagation. Over the past several years, numerous practical and theoretically sound elliptic location estimators have been developed to accomplish this task. Below is a brief summary of the relevant works.

When $\{ e_{m,l} \}$ follow zero-mean Gaussian distributions, least squares (LS)-based location estimators, whether linear or nonlinear, are employed to achieve statistical efficiency from the available BRs \cite{SMKay}. Linear LS approaches are perhaps the most extensively studied in the literature, thanks to the computational convenience they offer. By converting the nonlinear BR expressions into models linear in $\bm{x}$, these techniques allow for closed-form solutions \cite{MDianat,ANoroozi1,MEinemo,RAmiri4,ANoroozi2}, with some approaching closed form \cite{RAmiri1}. On the other hand, current progress in advanced iterative optimization solvers enables the direct implementation of nonlinear LS position estimators in a computationally efficient manner. Within the particular field of elliptic localization, solvers of this kind include Lagrange-type iterative algorithms based on the Lagrange programming neural network (LPNN) \cite{JLiang3} and the alternating direction method of multipliers (ADMM) \cite{JLiang2}. In addition, the constrained LS problems formulated for elliptic localization in Gaussian noise can be solved by means of semidefinite programming (SDP) \cite{RAmiri3,ZZheng}, a well-established convex optimization methodology.

More recently, there has been a growing number of investigations into elliptic localization in scenarios of greater practical significance, specifically in the presence of outlier measurements. The justification for such research lies in the fact that the daily operations of localization systems are often impacted by adverse environmental factors, including non-line-of-sight (NLOS) signal propagation, Byzantine faults, and interference \cite{IGuvenc,XMei3}. They will contribute to non-Gaussian $\{ e_{m,l} \}$, translate to outliers in $\{ \hat{r}_{m,l} \}$, and eventually degrade the positioning performance of conventional LS location estimators \cite{AMZoubir}.

Perhaps the most widely adopted strategy in the existing literature to handle outlying BRs is to exploit principles of robust $M$-estimation \cite{JLiang,ZShi2,ZYu,WXiongGRSL,WXiongTGRS,XZhao,WXionglp}. Approaches falling within this category operate by replacing the non-outlier-resistant $\ell_{2}$ loss in the LS formulation with cost functions that exhibit reduced sensitivity to deviating samples. In the pioneering research \cite{JLiang}, a correntropy-induced loss was introduced as the substitute for its $\ell_{2}$ counterpart, and a half-quadratic algorithm was devised to tackle the associated optimization problem. Later on, there have also been successful applications of more computationally lightweight iterative schemes to address the least absolute deviation (LAD)/$\ell_{1}$-minimization formulations for robust elliptic localization. These include the LPNN in \cite{ZShi2}, the iterative message passing (MP) algorithm in \cite{ZYu}, and the ADMM in \cite{WXiongGRSL,WXiongTGRS}. Expanding on the LPNN in \cite{ZShi2}, the authors of \cite{XZhao} developed an iteratively reweighted LAD (IRLAD) scheme tailored for $\ell_{p}$-minimization, particularly when $p=0$. The aim of this work was to enhance the robustness of the elliptic position estimator in specific outlier scenarios. In \cite{WXionglp}, majorization--minimization (MM) was incorporated into an iteratively reweighted LS (IRLS) framework to achieve statistical optimality in the case of stably distributed $\{ e_{m,l} \}$ through least $\ell_{p}$-norm estimation.

In our latest study \cite{WXiongCASTELO}, we tackled NLOS outliers in elliptic localization exclusively. We achieved this by integrating additional error variables into the conventional LS formulation, thereby leveraging the positivity of NLOS biases. The SDP was utilized as the optimization solver.

Alternative methods to ensure accuracy in adverse localization environments encompass exploiting the sparsity of outliers \cite{XMei3}, integrating optimal geometry analysis \cite{XMei1}, employing information fusion techniques \cite{XMei2}, and carrying out residual-based data selection \cite{WXiong5}, to name but a few.

This work is motivated by the exceptional outlier-resistance exhibited by the Chebyshev center positioning approach in several other range-based localization contexts \cite{XShi1,XShi2}. Such an approach sets itself apart from the outlier-handling strategies mentioned earlier by effectively taking advantage of the error upper bound, which is available during the calibration and commissioning stage of the localization system. Essentially, incorporating the error upper bound supplements the basic inputs of BR measurements and sensor positions with extra prior information, a feature not typically accounted for by conventional elliptic localization schemes in the literature. This supplementary information, when correctly employed, can result in improved accuracy in location estimation \cite{XShi1,XShi2}.

Our contribution of this work is threefold, as outlined below:

i) In contrast to the common trend of using $M$-estimation \cite{JLiang,ZShi2,ZYu,WXiongGRSL,WXiongTGRS,XZhao,WXionglp} or adjusting traditional LS formulations by integrating additional error variables \cite{WXiongCASTELO}, we opt for a different path to perform robust elliptic localization. Our formulation is rooted in the worst-case robust approximation criterion \cite{XShi1,XShi2,SBoyd1}, presenting itself as a min-max optimization problem. In essence, the maximum possible $\bm{x}$-estimation error under uncertainty specified by $\{ \hat{r}_{m,l} \}$ with bounded $\{ e_{m,l} \}$, is minimized, rather than a certain robust loss function as in the $M$-estimators. This, from a geometric perspective, corresponds to locating the Chebyshev center within the associated feasible set.

ii) Since determining the Chebyshev center poses an NP-hard problem except in certain special cases \cite{DWu}, we reshape the original min-max formulation into an SDP framework. In addition, we analyze the complexity for implementing the presented SDP-based approach.

iii) Simulation results are provided to validate our proposal and demonstrate its superior performance compared to several off-the-shelf elliptic localization methods.

\section{Problem statement and error modeling}

In this section, we formulate the robust location estimation problem by following the worst-case approximation criterion. Furthermore, we detail the measurement error modeling strategy under consideration.

In order to employ such a set-based, worst-case approach, it is necessary to first describe the uncertainty (to be maximized) across a set of possible values, namely, the feasible set \cite{SBoyd1}. To accomplish this, we make the additional assumption that the magnitude of the BR measurement errors $\{ e_{m,l} \}$ is upper bounded by some given constant $\rho > 0$:
\begin{equation}{\label{boundedness}}
	|e_{m,l}| \leq \rho.
\end{equation}
It is worth highlighting that similar assumptions have been commonly adopted in the literature concerning range-based localization methods, such as those utilizing the received signal strength \cite{STomic4}, time-of-arrival \cite{STomic4,GWang,XShi1}, and time-difference-of-arrival \cite{GWang2,XShi2} measurements. These requirements prove to be milder when compared to precisely characterizing the corresponding statistical distributions and, in practical terms, the error upper bound can be readily estimated during the calibration and commissioning stage using training data \cite{GWang2}.

Using $\bm{z} \in \mathbb{R}^{2}$ to denote a feasible target location estimate, we can now express the following min-max problem:
\begin{align}{\label{min_max}}
	\min_{\bm{x}}\max_{\bm{z} \in \mathcal{C}_{\bm{z}}}{\| \bm{x} - \bm{z} \|}_{2},
\end{align}
where
\begin{subequations}
\begin{align}
	\mathcal{C}_{\bm{z}} =& \big\{ \bm{z} \in \mathbb{R}^{H} \big| \underline{r}_{m,l} \leq {\| \bm{z} - \bm{t}_{m} \|}_{2} + {\| \bm{z} - \bm{s}_{l} \|}_{2} \leq \overline{r}_{m,l}, \nonumber\\
	&~m = 1,...,M,~l = 1,...,L \big\}, \label{Def_C}\\
	&\underline{r}_{m,l} = \hat{r}_{m,l} - \rho,~m = 1,...,M,~l = 1,...,L,\\
	&\overline{r}_{m,l} = \hat{r}_{m,l} + \rho,~m = 1,...,M,~l = 1,...,L.
\end{align}
\end{subequations}
Here, $\mathcal{C}_{\bm{z}}$ is the feasible set determined according to (\ref{meas_BR}) and (\ref{boundedness}), which amounts to the area where the ellipses defined by (\ref{Def_C}) intersect.

The geometric interpretation of (\ref{min_max}) is to identify the Chebyshev center nestled within $\mathcal{C}_{\bm{z}}$, viz., the center of the minimal-radius circle enclosing $\mathcal{C}_{\bm{z}}$.

In this contribution, we assume that $\{ e_{m,l} \}$ are independent and identically distributed and follow two-component Gaussian mixture impulsive distribution \cite{FYin1}:
\begin{align}{\label{pdf_GM}}
	&e = \beta \mathcal{N}(\mu,\sigma^2) + (1-\beta) \mathcal{N}(\check{\mu},\check{\sigma}^2),
\end{align}
where $e$ is the random variable being characterized and $\beta$ specifies the proportions in the two-component mixture. $\mathcal{N}$ denotes the standard Gaussian distribution. $\check{\mu}$ and $\check{\sigma}$ are the mean and standard deviation of the second Gaussian mixture component (to which an impulsive $e$ and an associated outlier measurement are attributed), respectively. In contrast, $\mu$ and $\sigma$ correspond to their counterparts associated with the first Gaussian mixture component.

The error modeling option (\ref{pdf_GM}) can offer certain advantages in the realm of localization \cite{FYin1}. This is owing to its compatibility and alignment with the superposition form commonly observed in measurement errors when they are introduced into location data acquired under NLOS conditions. The NLOS propagation of signals is a ubiquitous adverse environmental factor that touches on numerous positioning applications \cite{IGuvenc}. Such a superposition, typically, manifests itself as composite of a lower-level Gaussian disturbance component stemming from thermal fluctuations and a nonnegative random variable representing the NLOS bias \cite{IGuvenc,FYin1}.

\section{SDP formulation and complexity analysis}

While its structure is clear and comprehensible, dealing with the problem (\ref{min_max}) can be challenging due to the following two reasons. First, the task of determining the Chebyshev center is itself an NP-hard problem except in certain special cases \cite{DWu}. Second, the presence of multiple local minima and/or maxima makes the optimization process for min-max formulations more complex compared to many other standard optimization problems. In this section, instead of directly addressing (\ref{min_max}), we will cast it as an SDP problem that can be easier to solve.

As per the geometric interpretation described earlier, our objective is to locate the center of the smallest-radius circle containing the feasible set $\mathcal{C}_{\bm{z}}$. Based on this interpretation, we begin by transforming the intricate min-max formulation (\ref{min_max}) into the following epigraph-form expression:
\begin{align}{\label{min_max_reform}}
	\min_{\breve{\bm{x}},\bm{z},\bm{\gamma},t}t,~~\textup{s.t.}~\bm{z} = \breve{\bm{x}} + \bm{\gamma},~{\| \bm{\gamma} \|}_{2} \leq t,~\bm{z} \in \mathcal{C}_{\bm{z}},
\end{align}
where $\breve{\bm{x}} \in \mathbb{R}^{2}$ is the variable vector for the Chebyshev center of our interest, $\bm{z} = \breve{\bm{x}} + \bm{\gamma}$ defines a circle centered at $\breve{\bm{x}}$ with a radius ${\| \bm{\gamma} \|}_{2}$, $\bm{\gamma} \in \mathbb{R}^{2}$, and $t \in \mathbb{R}$. Rewriting the inequalities in (\ref{Def_C}) in quadratic form yields
\begin{subequations}
\begin{align}
	\underline{r}_{m,l}^2 + g_{m}^t - 2\underline{r}_{m,l}d_{m}^t &= (\underline{r}_{m,l} - d_{m}^t)^2 \leq g_{l}^s,\\
	\overline{r}_{m,l}^2 + g_{m}^t - 2\overline{r}_{m,l}d_{m}^t &= (\overline{r}_{m,l} - d_{m}^t)^2 \geq g_{l}^s,
\end{align}
\end{subequations}
where
\begin{align}{\label{cons_gmt_orig}}
	g_{m}^t &= (d_{m}^t)^2 = {\| \bm{z} - \bm{t}_{m} \|}_{2}^2
\end{align}
and
\begin{align}{\label{cons_gls_orig}}
	g_{l}^s &= {\| \bm{z} - \bm{s}_{l} \|}_{2}^2
\end{align}
represent the squared Euclidean distance between the $m$th transmitter and the target, and that between the target and the $l$th receiver, respectively. Similarly, we re-express ${\| \bm{\gamma} \|}_{2} \leq t$ as
\begin{equation}
	{\| \bm{\gamma} \|}_{2}^2 \leq t^2,~t \geq 0,
\end{equation}
which is equivalent to
\begin{subequations}{\label{cons_innercir}}
	\begin{align}
	\begin{bmatrix} t\bm{I}_{2} & \bm{\gamma} \\ \bm{\gamma}^T & t \end{bmatrix} &\succeq \bm{0}_{3 \times 3},\label{cons_innercir_1}\\
	t &\geq 0.\label{cons_innercir_2}
	\end{align}
\end{subequations}

With these reformulations, the constrained optimization problem (\ref{min_max_reform}) now reads
\begin{subequations}{\label{min_max_reform2}}
	\begin{align}
		&\min_{\breve{\bm{x}},\bm{z},\bm{\gamma},t,\{ d_{m}^t \},\{ g_{m}^t \},\{ g_{l}^s \},\lambda}t\nonumber\\
		\textup{s.t.}~&\textup{(\ref{cons_innercir})},\nonumber\\
		&\bm{z} = \breve{\bm{x}} + \bm{\gamma},\label{cons_zxgamma}\\
		&\underline{r}_{m,l}^2 + g_{m}^t - 2\underline{r}_{m,l}d_{m}^t \leq g_{l}^s,~\forall(m,l),\label{cons_ineq_quad_1}\\
		&\overline{r}_{m,l}^2 + g_{m}^t - 2\overline{r}_{m,l}d_{m}^t \geq g_{l}^s,~\forall(m,l),\label{cons_ineq_quad_2}\\
		&g_{m}^t = \begin{bmatrix} \bm{t}_{m} \\ -1 \end{bmatrix}^T \begin{bmatrix} \bm{I}_{2} & \bm{z} \\ \bm{z}^T & \lambda \end{bmatrix} \begin{bmatrix} \bm{t}_{m} \\ -1 \end{bmatrix},~m = 1,...,M,\label{cons_gmt}\\
		&g_{l}^s = \begin{bmatrix} \bm{s}_{l} \\ -1 \end{bmatrix}^T \begin{bmatrix} \bm{I}_{2} & \bm{z} \\ \bm{z}^T & \lambda \end{bmatrix} \begin{bmatrix} \bm{s}_{l} \\ -1 \end{bmatrix},~l = 1,...,L,\label{cons_gls}\\
		&\lambda = \bm{z}^T\bm{z},\label{cons_lambdaeq}\\
		&g_{m}^t = (d_{m}^t)^2,\label{cons_gmteq}
	\end{align}
\end{subequations}
where $\lambda \in \mathbb{R}$ is an additional variable representing the squared Euclidean distance for $\bm{z}$. The constraint (\ref{cons_zxgamma}) originates from (\ref{min_max_reform}), whereas the semidefinite cone constraints (\ref{cons_gmt}) and (\ref{cons_gls}) are directly derived from (\ref{cons_gmt_orig}) and (\ref{cons_gls_orig}), respectively.

The non-affine equality constraints (\ref{cons_lambdaeq}) and (\ref{cons_gmteq}) contribute to the nonconvexity of (\ref{min_max_reform2}). By discarding the rank-1 requirement \cite{ZQLuo} and relaxing them to
\begin{equation}{\label{cons_lambda}}
	\begin{bmatrix} \bm{I}_{2} & \bm{z} \\ \bm{z}^T & \lambda \end{bmatrix} \succeq \bm{0}_{3 \times 3}
\end{equation}
and
\begin{equation}{\label{cons_dmt}}
	\begin{bmatrix} 1 & d_{m}^t \\ d_{m}^t & g_{m}^t \end{bmatrix} \succeq \bm{0}_{2 \times 2},~m = 1,...,M,
\end{equation}
we obtain an SDP formulation:
\begin{align}{\label{Form_SDP}}
	&\min_{\breve{\bm{x}},\bm{z},\bm{\gamma},t,\{ d_{m}^t \},\{ g_{m}^t \},\{ g_{l}^s \},\lambda}t\nonumber\\
	\textup{s.t.}~&\textup{(\ref{cons_innercir})},~\textup{(\ref{cons_zxgamma})--(\ref{cons_gls})},~\textup{(\ref{cons_lambda})},~\textup{(\ref{cons_dmt})}.
\end{align}
Here, $\bm{A} \succeq \bm{0}_{a \times a}$ means that the $a \times a$ symmetric matrix $\bm{A}$ is positive semidefinite.

The SDP problem (\ref{Form_SDP}) falls into the domain of convex optimization and, therefore, can be easily solved using interior-point algorithms with a worst-case complexity on the order of \cite{GWang2}
\begin{align}{\label{Worst_Case_Complex}}
	\sqrt{\sum_{n=1}^{N_{\textup{SD}}}D_{\textup{SD,n}}}\!\left(\!N_{\textup{Var}}^2 \sum_{n=1}^{N_{\textup{SD}}} D_{\textup{SD,n}}^2\!+\!N_{\textup{Var}} \sum_{n=1}^{N_{\textup{SD}}} D_{\textup{SD,n}}^3\!+\!N_{\textup{Var}}^3\!\right)\!\log\big( \tfrac{1}{\nu} \big),
\end{align}
where $N_{\textup{SD}}$, $D_{\textup{SD,n}}$, $N_{\textup{Var}}$, and $\nu$ denote the number of semidefinite cone constraints, the dimension of the $n$th semidefinite cone constraint, the number of optimization variables, and the precision, respectively. It is evident that $N_{\textup{SD}} = 5+2ML+2M+L$ and $N_{\textup{Var}} = 8+2M+L$ in this particular case. Specifically, there are $3+2ML+M+L$ semidefinite cone constraints sized 1 (corresponding to the constraints in (\ref{cons_innercir_2}) and (\ref{cons_zxgamma})--(\ref{cons_gls}), despite some not being inequalities), $M$ semidefinite cone constraints sized 2 (corresponding to the constraints in (\ref{cons_dmt})), and 2 semidefinite cone constraints sized 3 (corresponding to the constraints in (\ref{cons_innercir_1}) and (\ref{cons_lambda})).

Now, let us compute each part of the expression in (\ref{Worst_Case_Complex}) as follows: 
\begin{subequations}
	\begin{align}
		\sqrt{\sum_{n=1}^{N_{\textup{SD}}}D_{\textup{SD,n}}} &= \sqrt{9+2ML+3M+L},\\
		N_{\textup{Var}}^2 \sum_{n=1}^{N_{\textup{SD}}} D_{\textup{SD,n}}^2 &= (8+2M+L)^2(21+2ML+5M+L),\\
		N_{\textup{Var}} \sum_{n=1}^{N_{\textup{SD}}} D_{\textup{SD,n}}^3 &= (8+2M+L)(57+2ML+9M+L),\\
		N_{\textup{Var}}^3 &= (8+2M+L)^3.
	\end{align}
\end{subequations}
For the first part, $\sqrt{9+2ML+3M+L}$, as $M$ and $L$ approach infinity, the dominant term becomes $2ML$, making $\mathcal{O}(\sqrt{9+2ML+3M+L}) = \mathcal{O}(\sqrt{ML})$. For the second part, $(8+2M+L)^2(21+2ML+5M+L)$, as $M$ and $L$ tend to infinity, the leading term becomes $(2ML)^{2}$, resulting in $\mathcal{O}((8+2M+L)^2(21+2ML+5M+L)) = \mathcal{O}(M^2L^2)$. For the third part, $(8+2M+L)(57+2ML+9M+L)$, as $M$ and $L$ increase without bound, the dominant term turns out to be $(2M+L)(2ML)$, making it $\mathcal{O}((M+L)ML)$. Finally, considering the fourth part, $(8+2M+L)^3$, we can ignore the constant 8 in the limit as $M$ and $L$ become large. Then, when cubed, the cross products sum the powers of $M$ and $L$ to 3, indicating that $N_{\textup{Var}}^3$ is insignificant compared to $N_{\textup{Var}}^2 \sum_{n=1}^{N_{\textup{SD}}} D_{\textup{SD,n}}^2$ in terms of limit computation. Plugging these terms into (\ref{Worst_Case_Complex}) reveals that the estimated worst-case complexity for solving (\ref{Form_SDP}) using interior-point methods is $\mathcal{O}((ML)^{2.5})$.

In order to illustrate the scale of this complexity, we include Table \ref{Table_1}, where the complexities of several existing elliptic localization methods capable of handling outliers are showcased. These approaches include (i) the LAD estimator in \cite{ZShi2}, implemented via LPNN, (ii) the LAD estimator in \cite{ZYu}, implemented via iterative MP, (iii) the $\ell_{p}$ estimator with $p=0$ in \cite{XZhao}, implemented via LPNN-based IRLAD, (iv) the $\ell_{p}$ estimator with $p=1.5$ in \cite{WXionglp}, implemented via MM-based IRLS, and (v) the convex approximation based solution \texttt{CASTELO} \cite{WXiongCASTELO}, which is a transformed LS estimator that incorporates additional bias-representing estimation variables. They will be contrasted with our proposed SDP-based technique (termed \texttt{min-max}) in the simulations in Section \ref{Sect_NR}. In Table \ref{Table_1}, $N_{\textup{LPNN}}$, $N_{\textup{MP}}$, $N_{\textup{IRLAD}}$, $N_{\textup{IRLS}}$, and $N_{\textup{MM}}$ represent the discrete realization steps for LPNN (typically several thousands), the number of MP iterations (usually several tens), the number of IRLAD iterations (usually several tens), the number of IRLS iterations (usually several tens), and the number of MM iterations (usually a few), respectively. It is clear that \texttt{min-max} remains computationally competitive in typical elliptic localization configurations, where the values of $M$ and $L$ are not very large in magnitude in most cases \cite{LRui1}.

\begin{table*}[!t]
	\renewcommand{\arraystretch}{1.2}
	\caption{Complexities of several existing elliptic localization methods capable of handling outliers and our proposal}
	\label{Table_1}
	\centering
	\begin{tabular}{|c|c|}
		\hline
		\textbf{Method} & \textbf{Complexity} \\
		\hline
		\texttt{$\ell_{1}$-LPNN} \cite{ZShi2} & $\mathcal{O}(N_{\textup{LPNN}}(M+L))$ \\
		\hline
		\texttt{MP} \cite{ZYu} & $\mathcal{O}(N_{\textup{MP}}(ML))$ \\
		\hline
		\texttt{$\ell_{0}$-IIRW} \cite{XZhao} & $\mathcal{O}(N_{\textup{IRLAD}}N_{\textup{LPNN}}(M+L))$ \\
		\hline
		\texttt{$\ell_{1.5}$-IRLS} \cite{WXionglp} & $\mathcal{O}(N_{\textup{IRLS}}N_{\textup{MM}}(ML))$ \\
		\hline
		\texttt{CASTELO} \cite{WXiongCASTELO} & $\mathcal{O}((ML)^{3.5})$ \\
		\hline
		\texttt{min-max} (proposed) & $\mathcal{O}((ML)^{2.5})$ \\
		\hline
	\end{tabular}
\end{table*}

\section{Numerical results}

\label{Sect_NR}

In this section, we conduct numerical tests on a laptop equipped with a 2.8 GHz CPU and 8 GB of RAM to evaluate the performance of \texttt{min-max} (code available at \url{https://github.com/w-x-xiong/min-max}). For simulation purposes, we employ a distributed MIMO radar system comprising $M=3$ transmitters and $L=4$ receivers. We assume a deterministic deployment of the target and transmit/receive antennas, with their positions specified as follows: $\bm{t}_{1} = [-200, -300]^T$ m, $\bm{t}_{2} = [-200, 300]^T$ m, $\bm{t}_{3} = [200, 300]^T$ m, $\bm{s}_{1} = [-450, -450]^T$ m, $\bm{s}_{2} = [450, 450]^T$ m, $\bm{s}_{3} = [0, 600]^T$ m, $\bm{s}_{4} = [600, 0]^T$ m, and $\bm{x} = [100, 100]^T$ m. We assess the positioning accuracy of \texttt{min-max} by comparing its root-mean-square error (RMSE):
\begin{equation}{\label{def_RMSE}}
	\textup{RMSE} = \sqrt{\tfrac{1}{N_{\textup{MC}}}\sum_{n=1}^{N_{\textup{MC}}}{\| \tilde{\bm{x}}^{\{ n \}} - \bm{x} \|}_{2}^2}
\end{equation}
to that of the approaches from Table \ref{Table_1}. In addition, we take into account the following two LS-based elliptic localization schemes: a) the constrained weighted LS estimator presented in \cite{RAmiri1}, allowing for an exact solution (termed \texttt{exact}) and b) the nonlinear LS estimator directly implemented through the LPNN \cite{JLiang3} (termed \texttt{$\ell_{2}$-LPNN}). In (\ref{def_RMSE}), $\tilde{\bm{x}}^{\{ n \}}$ represents the target location estimate in the $n$th Monte Carlo (MC) run. $N_{\textup{MC}}$ denotes the number of ensemble MC runs, which is fixed at 100 in our simulations. $\{ e_{m,l} \}$ are generated based on the Gaussian mixture impulsive noise modeling strategy in (\ref{pdf_GM}). Regarding the implementation of \texttt{min-max}, we solve the associated SDP problem (\ref{Form_SDP}) using the MATLAB toolbox CVX \cite{MGrant}, with the SDPT3 solver. Unless stated otherwise, the error upper bound required by \texttt{min-max} is exactly provided as $\rho = \max_{(m,l) \in \{ \{ 1,...,M \} \times \{ 1,...,L \} \} }|\hat{r}_{m,l} - {\| \bm{x} - \bm{t}_{m} \|}_{2} - {\| \bm{x} - \bm{s}_{l} \|}_{2}|$.

\begin{figure}[!t]
	\centering
	\includegraphics[width=6.5in]{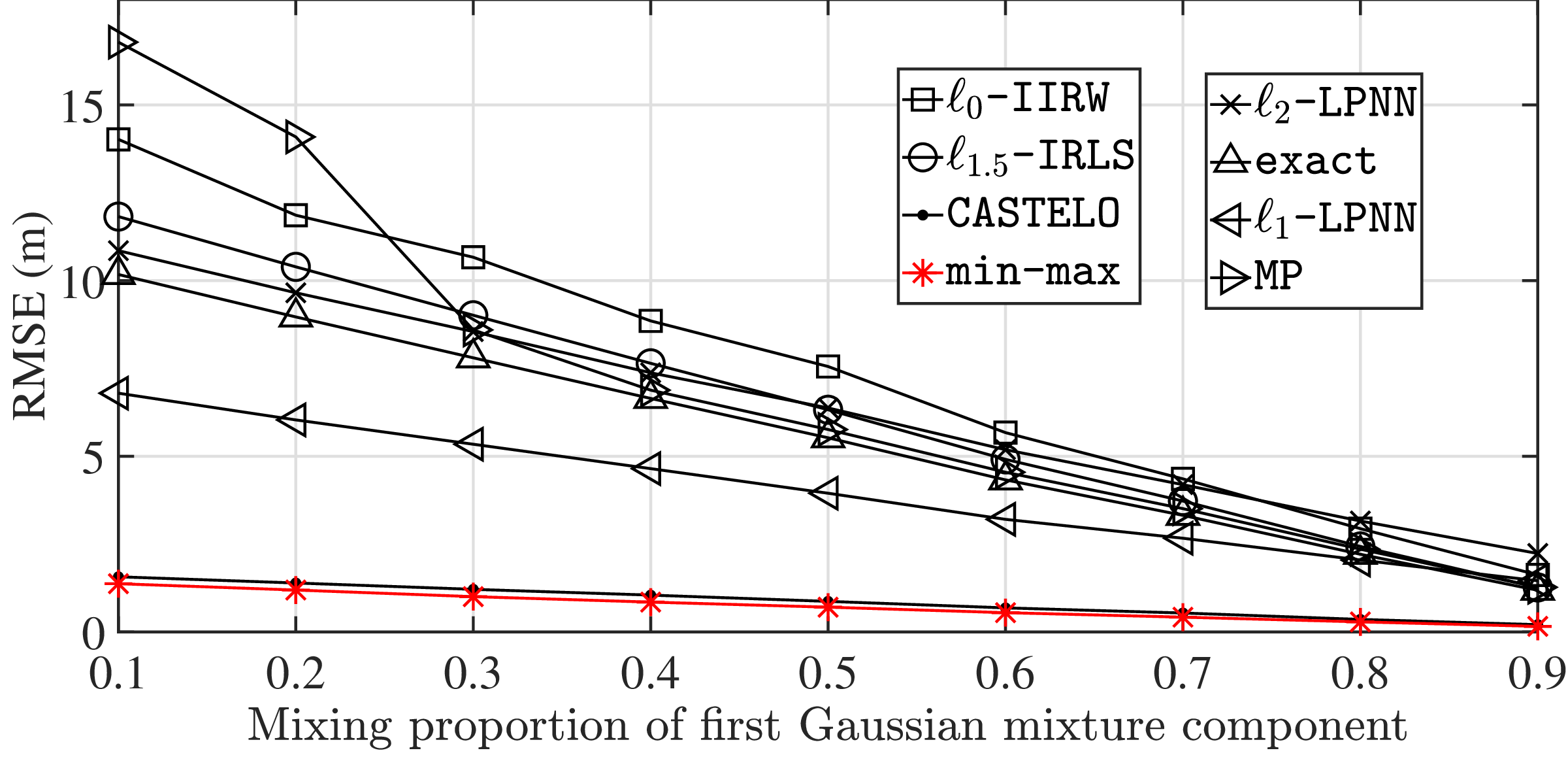}
	\caption{RMSE versus $\beta$ when $\mu=0$, $\sigma=1$ m, $\check{\mu}=20$ m, and $\check{\sigma}=1$ m.}
	\label{Fig_RMSEvsprop}
\end{figure}

In our first experiment, we vary the proportion of the first Gaussian mixture component, $\beta$, from 0.1 to 0.9, while keeping other parameters in (\ref{pdf_GM}) fixed: $\mu=0$, $\sigma=1$ m, $\check{\mu}=20$ m, and $\check{\sigma}=1$ m. Parameter setups with a progressively increasing $\beta$, as indicated by (\ref{pdf_GM}), give rise to scenarios in which the likelihood of NLOS conditions gradually decreases from high to low probability. Fig. \ref{Fig_RMSEvsprop} depicts the corresponding RMSE results. We observe that \texttt{min-max} slightly surpasses the current state-of-the-art (SOTA) method \texttt{CASTELO} and consistently outperforms the other approaches, providing the highest positioning accuracy across the entire $\beta$ range under investigation. This can be attributed to the effective utilization of the error upper bound $\rho$ as supplementary information, which helps constrain the feasible set for convex optimization, leading to significantly enhanced estimation accuracy. It is important to note that our aim is not to create an unfair advantage in our numerical comparisons, but rather to highlight the effectiveness of leveraging supplementary prior information, particularly the error bound. As such techniques are not yet prevalent in the existing literature on elliptic localization, here we compare our approach against methods that do not incorporate $\rho$. Also apparent from Fig. \ref{Fig_RMSEvsprop} is the improvement in the performance of all estimators as $\beta$ increases. This is reasonable because a higher $\beta$ signifies a greater likelihood of line-of-sight (LOS), implying that the algorithms are less affected by NLOS errors and can place more trust in BR measurements that are more dependable. The state-of-the-art scheme \texttt{CASTELO} can achieve results nearly comparable to \texttt{min-max} without utilizing the error upper bound. This is because: (i) \texttt{CASTELO} was specifically designed to handle NLOS propagation situations by making use of the positivity of NLOS biases, and (ii) \texttt{CASTELO} introduces additional second-order cone (SOC) constraints, resulting in a tighter convex relaxation compared to relying solely on semidefinite relaxation. In contrast, our proposed approach, \texttt{min-max}, does not incorporate these two designs due to concerns that: (i) in broader outlier situations, the bias errors in the outlier BR measurements may not necessarily exhibit such positivity properties, and (ii) the inclusion of SOC constraints may lead to challenging convex hull issues, where the estimated target position must lie within the convex hull formed by transmitters, which may not always hold true in various localization configurations. Therefore, our proposal \texttt{min-max} can in fact be less prone to errors in practical use cases. \texttt{$\ell_{1}$-LPNN} emerges as the third-best solution for $\beta \leq 0.8$, affirming the considerable outlier-robustness of the $\ell_{1}$-minimization formulation and the LPNN's effectiveness as a reliable optimization solution in this context. These observations align closely with the findings from several existing works \cite{ZShi2,WXionglp,WXiongTGRS}.

\begin{figure}[!t]
	\centering
	\includegraphics[width=6.5in]{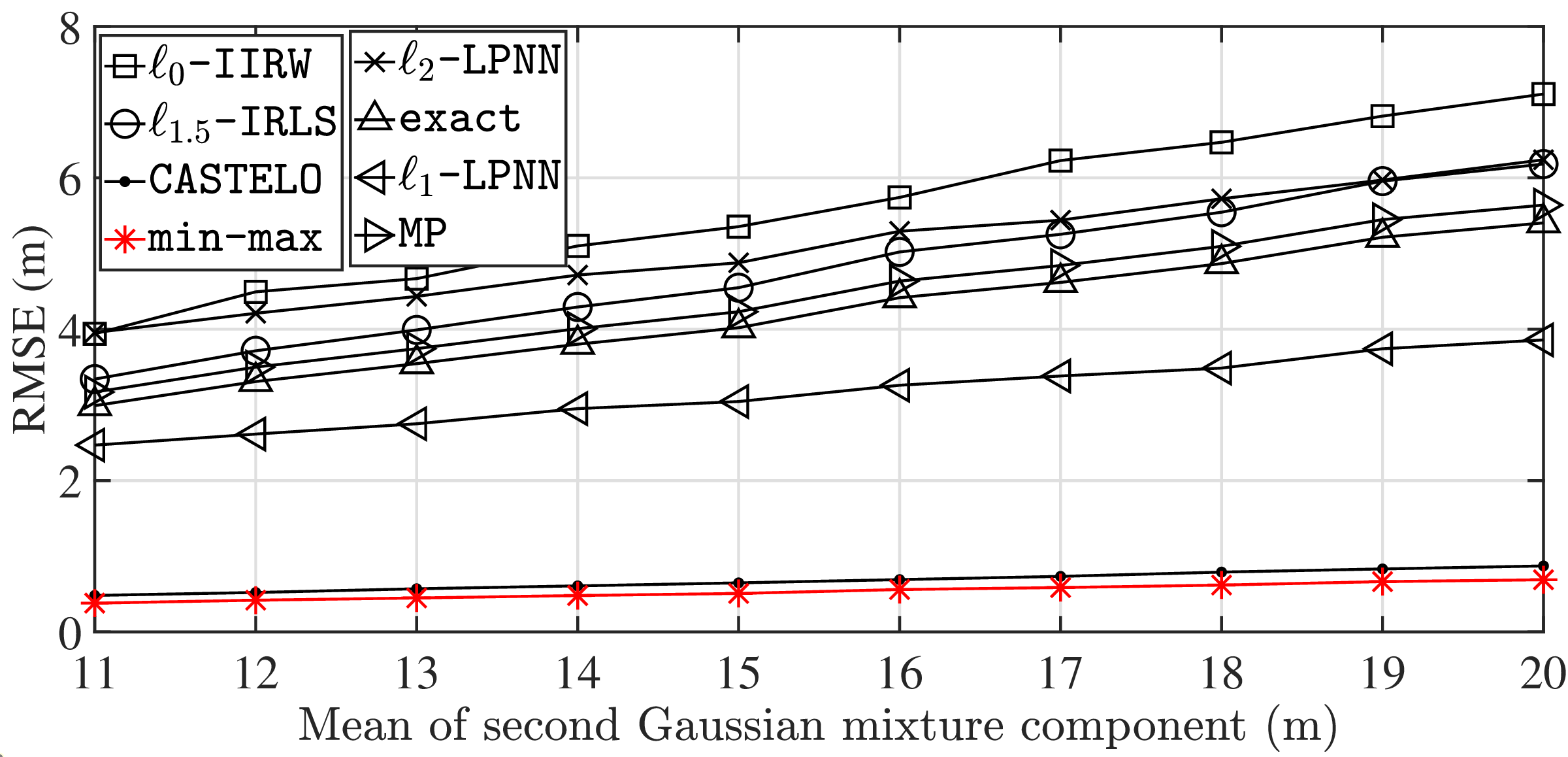}
	\caption{RMSE versus $\check{\mu}$ when $\beta=0.5$, $\mu=0$, $\sigma=1$ m, and $\check{\sigma}=1$ m.} 
	\label{Fig_RMSEvsmean2ndGM}
\end{figure}

\begin{figure}[!t]
	\centering
	\includegraphics[width=6.5in]{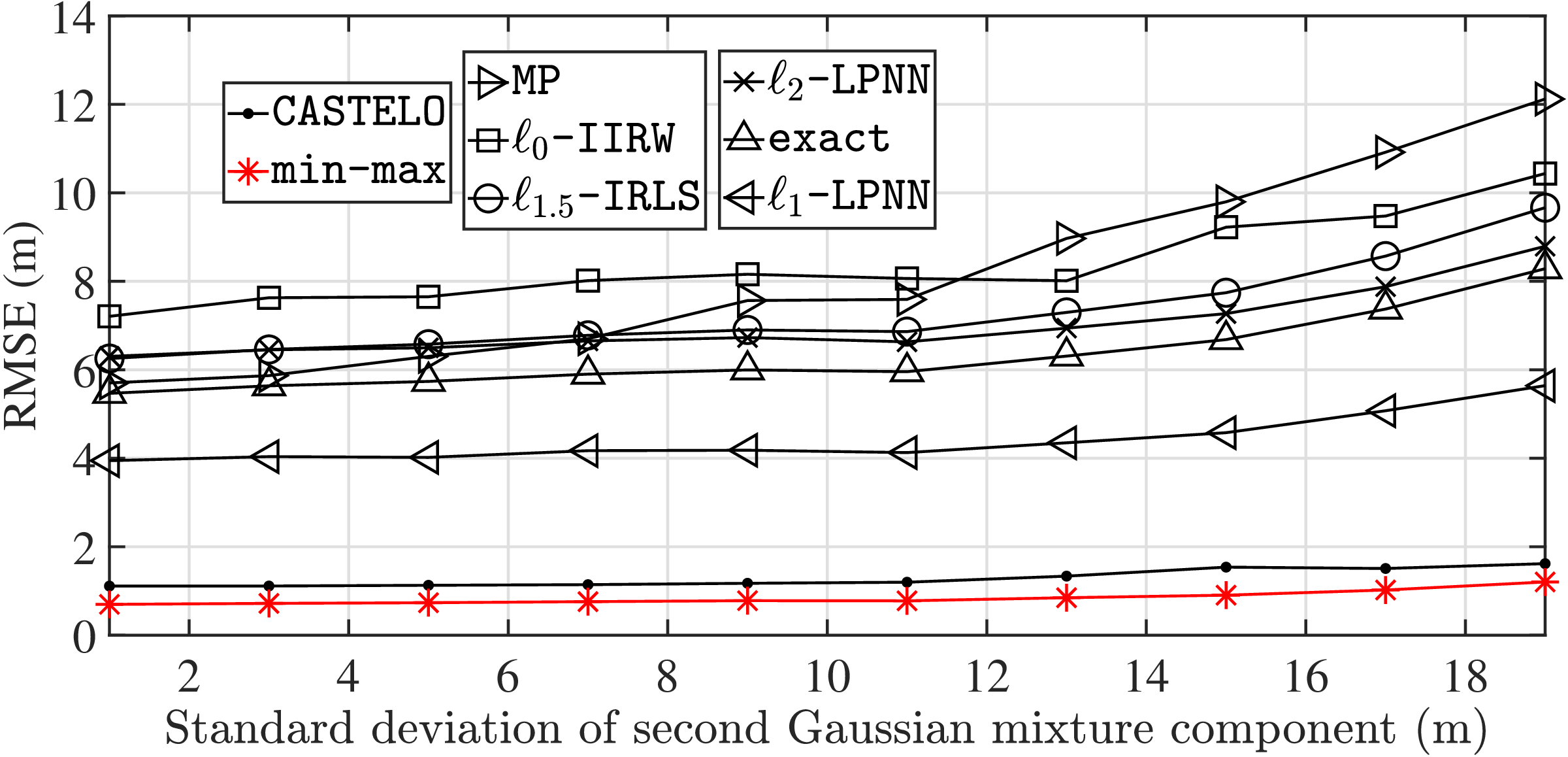}
	\caption{RMSE versus $\check{\sigma}$ when $\beta=0.5$, $\mu=0$, $\sigma=1$ m, and $\check{\mu}=20$ m.} 
	\label{Fig_RMSEvsstd2ndGM}
\end{figure}

Fig. \ref{Fig_RMSEvsmean2ndGM} illustrates how the RMSE varies with the mean of the second Gaussian mixture component, $\check{\mu}$, limited to the interval $[11, 20]$ m. These examinations are carried out under the fixed parameters of $\beta=0.5$, $\mu=0$, $\sigma=1$ m, and $\check{\sigma}=1$ m. Such a configuration defines test scenarios where LOS links distinctly stand out from their NLOS counterparts. The setting of $\beta = 0.5$ balances the probabilities of encountering LOS and NLOS conditions equally, ensuring a 50/50 split for a specific link. In Fig. \ref{Fig_RMSEvsmean2ndGM}, there is a clear hierarchy in RMSE values among the seven estimators within the specific $\check{\mu}$-region we are investigating: our proposed method \texttt{min-max} is the lowest, followed by \texttt{CASTELO}, \texttt{$\ell_{1}$-LPNN}, \texttt{exact}, \texttt{MP}, \texttt{$\ell_{1.5}$-IRLS}, \texttt{$\ell_{2}$-LPNN}, and \texttt{$\ell_{0}$-IIRW}, in that order. While our previous discussions regarding the superiority of \texttt{min-max} as well as the strength of \texttt{CASTELO} and \texttt{$\ell_{1}$-LPNN} still hold true for this test scenario, the overall inferiority of \texttt{$\ell_{0}$-IIRW} in this case may be due to the inherent unsuitability of $\ell_{0}$-minimization when outliers are not particularly sparse or spike-like \cite{WJZeng}. The RMSE values of the \texttt{$\ell_{1.5}$-IRLS} estimator fall between those of the \texttt{$\ell_{2}$-LPNN} estimator and the \texttt{$\ell_{1}$-LPNN} estimator, which is expected given that $1 < 1.5 < 2$. With an increase in $\check{\mu}$, indicating greater differences in error magnitudes between measurements associated with LOS- and NLOS-dominant links, the RMSE exhibits a consistent upward trend across all schemes. Subsequently, in Fig. \ref{Fig_RMSEvsstd2ndGM}, the results of RMSE versus the standard deviation of the second Gaussian mixture component, $\check{\sigma}$, are depicted under the specified conditions of $\beta=0.5$, $\mu=0$, $\sigma=1$ m, and $\check{\mu}=20$ m. For each approach, when the $\check{\sigma}$ value exceeds a certain level, the trend of RMSE rising with increasing $\check{\sigma}$/greater disparities between the dispersions of the two Gaussian mixture components becomes discernible. Again, \texttt{min-max} demonstrates a performance advantage in RMSE compared to others across the entire parameter range we are examining.

\begin{figure}[!t]
	\centering
	\includegraphics[width=6.5in]{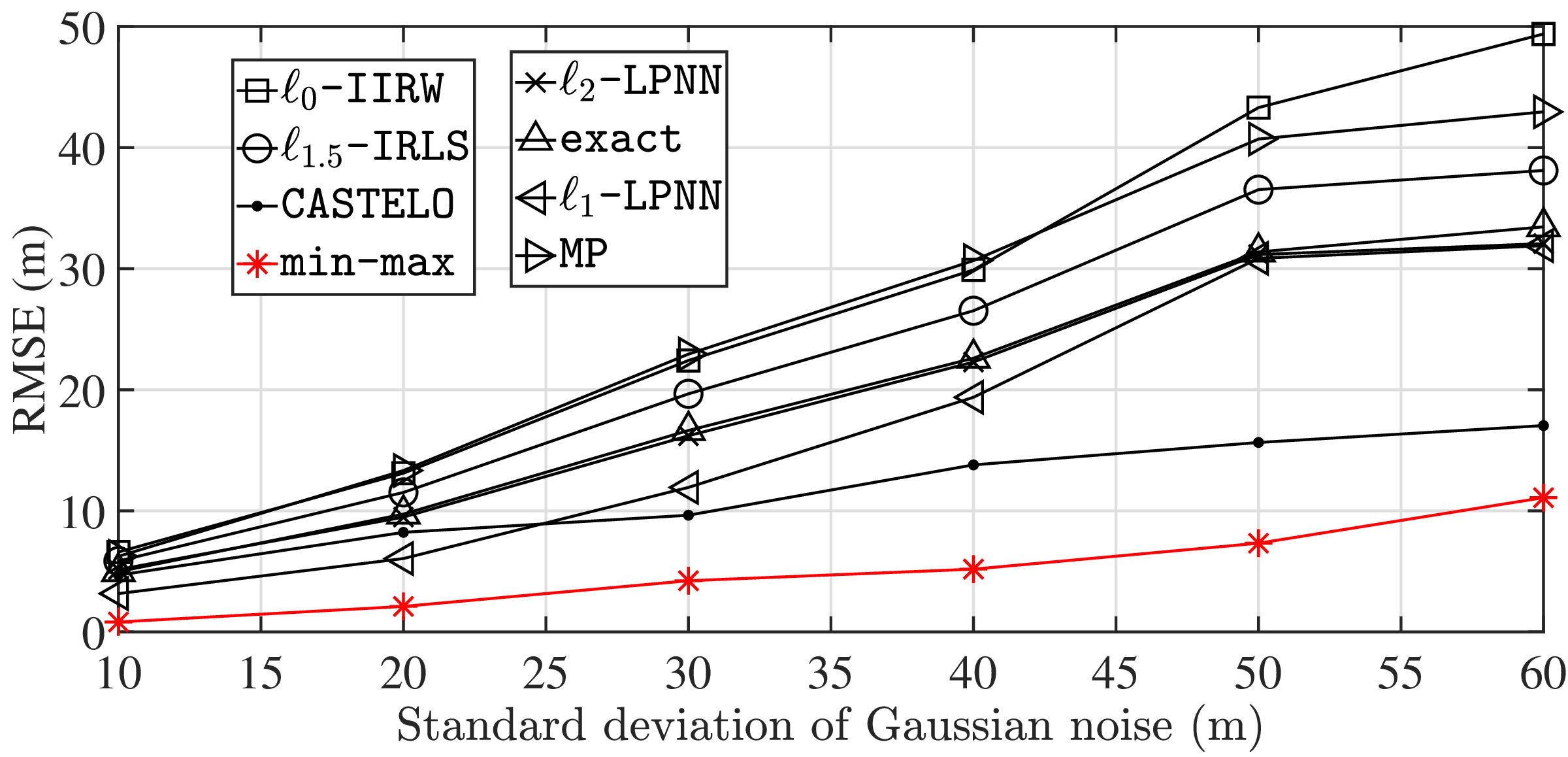}
	\caption{RMSE versus $\sigma$ when $\beta=1$ and $\mu=0$.} 
	\label{Fig_RMSEvsstdG}
\end{figure}

While the above experiments focused on elliptic localization in NLOS environments, there is also a research curiosity in assessing how well our proposed technique functions in noise situations characterized solely by zero-mean Gaussian distributions (i.e., where the adverse environmental factors are absent). To delve into this, Fig. \ref{Fig_RMSEvsstdG} plots the RMSE versus the standard deviation of the first Gaussian mixture component, $\sigma$, with $\beta$ set to 1 and $\mu$ at 0. We observe that \texttt{min-max} continues to stand out as the optimal solution in terms of RMSE, whereas \texttt{CASTELO} remains the second best. Moreover, the RMSE of both \texttt{min-max} and \texttt{CASTELO} increases at a similar rate as $\sigma$ grows, and this rate is much lower than that of the other estimators. This indicates their superior resistance to rising levels of Gaussian noise.

\begin{figure}[!t]
	\centering
	\includegraphics[width=6.5in]{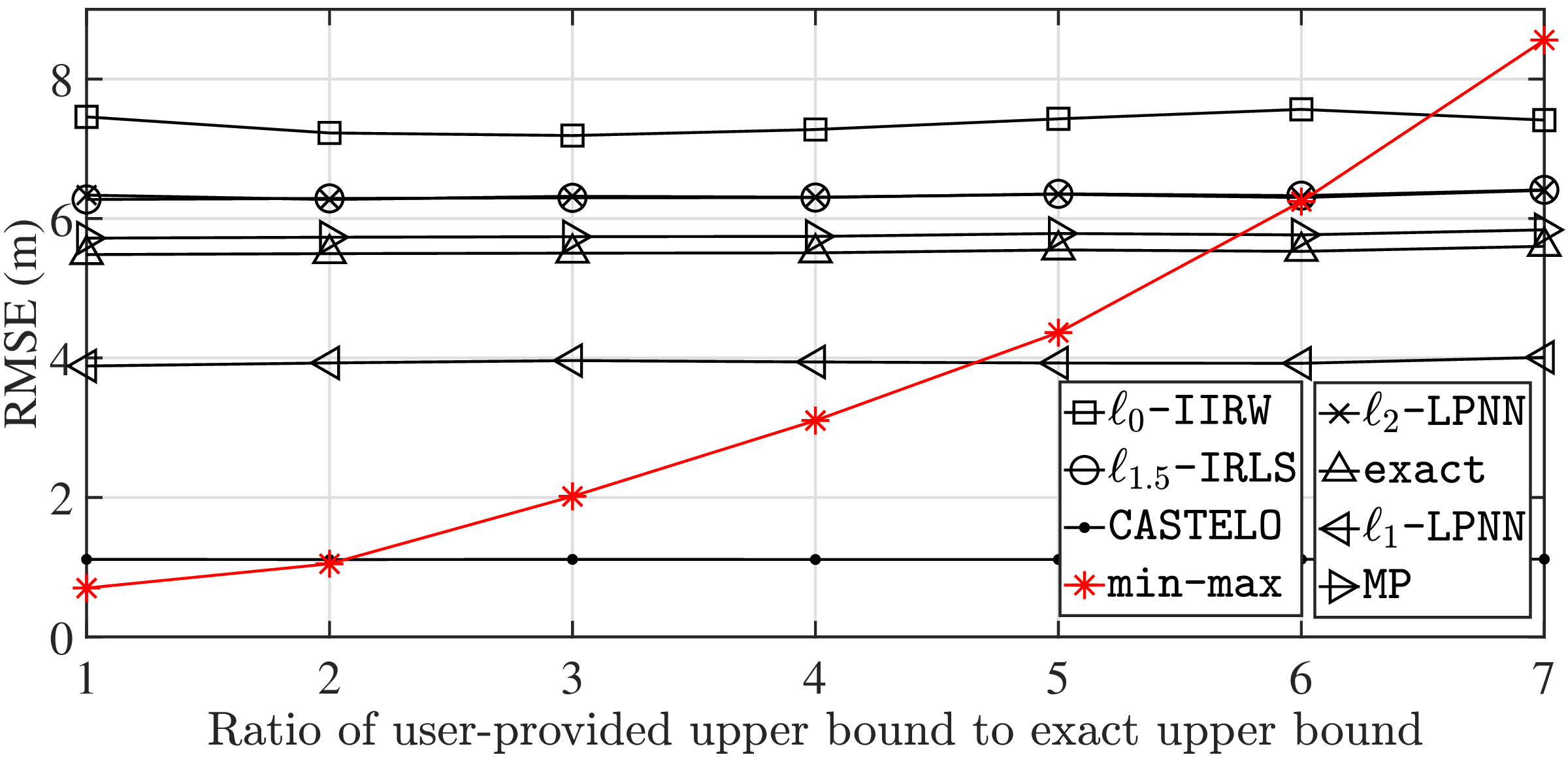}
	\caption{RMSE versus $\hat{\rho}/\rho$ when $\beta=0.5$, $\mu=0$, $\sigma=1$ m, $\check{\mu}=20$ m, and $\check{\sigma}=1$ m.} 
	\label{Fig_RMSEvsratio}
\end{figure}

Indeed, incorporating the prior information of the error upper bound provides an advantage similar to having an ``oracle'' that somehow informs us about the correctness of BR measurements. Different from the ideal scenarios previously considered, where \texttt{min-max} operated with the exact error upper bound $\rho = \max_{(m,l) \in \{ \{ 1,...,M \} \times \{ 1,...,L \} \} }|\hat{r}_{m,l} - {\| \bm{x} - \bm{t}_{m} \|}_{2} - {\| \bm{x} - \bm{s}_{l} \|}_{2}|$, we now investigate \texttt{min-max}'s sensitivity to the exactness of the user-provided upper bound, $\hat{\rho}$, by plotting the RMSE in relation to $\hat{\rho}/\rho$. The error-characterizing parameters in (\ref{pdf_GM}) are set as $\beta=0.5$, $\mu=0$, $\sigma=1$ m, $\check{\mu}=20$ m, and $\check{\sigma}=1$ m, and Fig. \ref{Fig_RMSEvsratio} plots the RMSE as $\hat{\rho}/\rho$ gradually increases from 1 to 7. The results demonstrated that as long as the user-provided upper bound is within twice the exact value, our approach maintains SOTA positioning accuracy. Even with a much looser upper bound---when the user-provided value exceeds twice the exact value---as long as it remains less than four times the exact value, our approach still ranks as the second-best solution. This suggests that the proposed method offers substantial flexibility when incorporating such supplementary prior information.

\section{Conclusion and future work}
\label{Sect_Cc}

Outliers may severely affect the performance of traditional LS-based localization schemes, making their location estimates less reliable. This can lead to serious problems, such as guiding navigation users to wrong destinations or presenting irrelevant suggestions in location-based services. Elliptic localization, a newly emerging target positioning technique spurred by the expansion of multistatic systems, also faces challenges with outlier conditions that require effective handling. In this short communication, we have presented a new formulation for elliptic localization with outliers based on the worst-case robust approximation criterion. Geometrically speaking, our goal was to locate the Chebyshev center given a feasible set defined according to (i) the known sensor positions, (ii) the available BR measurements, and (iii) the maximum limit for the magnitude of BR measurement errors. Unlike most existing strategies, which often utilize concepts like $M$-estimation or integrate error variables into conventional LS formulations, our method incorporates supplementary information on the error upper bound, typically accessible during the calibration stage of the localization solution using training data. It assists in creating a more precise feasible set, resulting in enhanced accuracy. The resulting problem, initially, posed itself as a min-max optimization task. Instead of directly addressing this challenging min-max formulation, we conducted a transformation based on its geometric interpretation and shifted our focus to an SDP approach that holds greater practical significance. Simulation results affirm the robustness of our proposed elliptic localization method to outliers, demonstrating an overall improvement compared to a number of existing solutions. Possible future research directions include (i) exploring alternative optimization solvers other than SDP to reduce computational complexity, and (ii) broadening the application of our method to more practical use cases of elliptic localization, such as scenarios where clock synchronization and/or transmitter positions are unavailable.



\begin{thebibliography}{99}  
	




	\bibitem{LRui1}
	L. Rui and K. C. Ho, ``Elliptic localization: Performance study and optimum receiver placement,'' \textit{IEEE Trans. Signal Process.}, vol. 62, no. 18, pp. 4673--4688, Sep. 2014.



	\bibitem{HCSo}
	H. C. So, ``Source localization: Algorithms and analysis,'' in \textit{Handbook of Position Location: Theory, Practice and Advances}, 2nd ed., S. A. Zekavat and M. Buehrer, Eds. New York, NY, USA: Wiley-IEEE Press, 2019, pp. 59--106.





	\bibitem{SMKay}
	S. M. Kay, \textit{Fundamentals of Statistical Signal Processing: Estimation Theory}, vol. 2. Cliffs, NJ, USA: Prentice-Hall, 1998.

	

	\bibitem{MDianat}
	M. Dianat, M. R. Taban, J. Dianat, and V. Sedighi, ``Target localization using least squares estimation for MIMO radars with widely separated antennas,'' \textit{IEEE Trans. Aerosp. Electron. Syst.}, vol. 49, no. 4, pp. 2730--2741, Oct. 2013.



	\bibitem{ANoroozi1}
	A. Noroozi and M. A. Sebt, ``Target localization from bistatic range measurements in multi-transmitter multi-receiver passive radar,'' \textit{IEEE Signal Process. Lett.}, vol. 22, no. 12, pp. 2445--2449, Dec. 2015.



	\bibitem{MEinemo}
	M. Einemo and H. C. So, ``Weighted least squares algorithm for target localization in distributed MIMO radar,'' \textit{Signal Process.}, vol. 115, pp. 144--150, Oct. 2015.




	\bibitem{RAmiri4}
	R. Amiri, F. Behnia, and H. Zamani, ``Asymptotically efficient target localization from bistatic range measurements in distributed MIMO radars,'' \textit{IEEE Signal Process. Lett.}, vol. 24, no. 3, pp. 299--303, Mar. 2017.

	


	\bibitem{ANoroozi2}
	A. Noroozi, M. A. Sebt, S. M. Hosseini, R. Amiri, and M. M. Nayebi, ``Closed-form solution for elliptic localization in distributed MIMO radar systems with minimum number of sensors,'' \textit{IEEE Trans. Aerosp. Electron. Syst.}, vol. 56, no. 4, pp. 3123--3133, Aug. 2020.
	

	
	\bibitem{RAmiri1}
	R. Amiri, F. Behnia, and M. A. M. Sadr, ``Exact solution for elliptic localization in distributed MIMO radar systems,'' \textit{IEEE Trans. Veh. Technol.}, vol. 67, no. 2, pp. 1075--1086, 2018.





	\bibitem{JLiang3}
	J. Liang, C. S. Leung, and H. C. So, ``Lagrange programming neural network approach for target localization in distributed MIMO radar,'' \textit{IEEE Trans. Signal Process.}, vol. 64, no. 6, pp. 1574--1585, Mar. 2016.



	\bibitem{JLiang2}
	J. Liang, Y. Chen, H. C. So, and Y. Jing, ``Circular/hyperbolic/elliptic localization via Euclidean norm elimination,'' \textit{Signal Process.}, vol. 148, pp. 102--113, Jul. 2018.



	\bibitem{RAmiri3}
	R. Amiri, F. Behnia, and M. A. M. Sadr, ``Positioning in MIMO radars based on constrained least squares estimation,'' \textit{IEEE Commun. Lett.}, vol. 21, no. 10, pp. 2222--2225, Oct. 2017.


	\bibitem{ZZheng}
	Z. Zheng, H. Zhang, W.-Q. Wang, ``Target localization in distributed MIMO radars via improved semidefinite relaxation,'' \textit{J. Franklin Inst.}, vol. 358, no. 10, pp. 5588--5598, Jul. 2021.



	\bibitem{IGuvenc}
	I. Guvenc and C.-C. Chong, ``A survey on TOA based wireless localization and NLOS mitigation techniques,'' \textit{IEEE Commun. Surveys Tuts.}, vol. 11, no. 3, pp. 107--124, Aug. 2009.



	\bibitem{XMei3}
	X. Mei, H. Wu, J. Xian, and B. Chen, ``RSS-based Byzantine fault-tolerant localization algorithm under NLOS environment,'' \textit{IEEE Commun. Lett.}, vol. 25, no. 2, pp. 474--478, Feb. 2021.



	\bibitem{AMZoubir}
	A. M. Zoubir, V. Koivunen, E. Ollila, and M. Muma, \textit{Robust Statistics for Signal Processing}. Cambridge, U.K.: Cambridge Univ. Press, 2018.



	\bibitem{JLiang}
	J. Liang, D. Wang, L. Su, B. Chen, H. Chen, and H. C. So, ``Robust MIMO radar target localization via nonconvex optimization,'' \textit{Signal Process.}, vol. 122, pp. 33--38, May 2016.



	\bibitem{ZShi2}
	Z. Shi, H. Wang, C. S. Leung, and H. C. So, ``Robust MIMO radar target localization based on Lagrange programming neural network,'' \textit{Signal Process.}, vol. 174, 107574, Sep. 2020.


	\bibitem{ZYu}
	Z. Yu, J. Li, Q. Guo, and T. Sun, ``Message passing based robust target localization in distributed MIMO radars in the presence of outliers,'' \textit{IEEE Signal Process. Lett.}, vol. 27, pp. 2168--2172, 2020.




	\bibitem{WXiongGRSL}
	W. Xiong, ``Denoising of bistatic ranges for elliptic positioning,'' \textit{IEEE Geosci. Remote Sens. Lett.}, vol. 20, pp. 1--3, 2023, Art. no. 3500503.



	\bibitem{WXiongTGRS}
	W. Xiong, G. Cheng, C. Schindelhauer, and H. C. So, ``Robust matrix completion for elliptic positioning in the presence of outliers and missing data,'' \textit{IEEE Trans. Geosci. Remote Sens.}, vol. 61, pp. 1--12, 2023, Art no. 5105912.



	\bibitem{XZhao}
	X. Zhao, J. Li, and Q. Guo, ``Robust target localization in distributed MIMO radar with nonconvex $\ell_{p}$ minimization and iterative reweighting,'' \textit{IEEE Commun. Lett.}, vol. 27, no. 12, pp. 3230--3234, Dec. 2023.



	\bibitem{WXionglp}
	W. Xiong, J. He, H. C. So, J. Liang, C.-S. Leung, ``$\ell_{p}$-norm minimization for outlier-resistant elliptic positioning in $\alpha$-stable impulsive interference,'' \textit{J. Franklin Inst.}, vol. 361, no. 1, pp. 21--31, Jan. 2024.



	\bibitem{WXiongCASTELO}
	W. Xiong, Z.-L. Shi, H. C. So, J. Liang, and Z. Wang, ``CASTELO: Convex approximation based solution to elliptic localization with outliers,'' \textit{Signal Process.}, vol. 218, 109380, May 2024.


	\bibitem{XMei1}
	X. Mei, D. Han, N. Saeed, H. Wu, C.-C. Chang, B. Han, T. Ma, and J. Xian, ``Trajectory optimization of autonomous surface vehicles with outliers for underwater target localization,'' \textit{Remote Sens.}, vol. 14, p. 4343, Sep. 2022.


	\bibitem{XMei2}
	X. Mei, D. Han, Y. Chen, H. Wu, and T. Ma, ``Target localization using information fusion in WSNs-based marine search and rescue,'' \textit{Alexandria Eng. J.}, vol. 68, pp. 227--238, Apr. 2023.



	\bibitem{WXiong5}
	W. Xiong, J. Bordoy, C. Schindelhauer, A. Gabbrielli, G. Fischer, D. J. Schott, F. Hoeflinger, S. J. Rupitsch, and H. C. So, ``Data-selective least squares methods for elliptic localization with NLOS mitigation,'' \textit{IEEE Sensors Lett.}, vol. 5, no. 7, pp. 1--4, Jul. 2021.



	\bibitem{XShi1}
	X. Shi, G. Mao, B. D. O. Anderson, Z. Yang, and J. Chen, ``Robust localization using range measurements with unknown and bounded errors,'' \textit{IEEE Trans. Wireless Commun.}, vol. 16, no. 6, pp. 4065--4078, Jun. 2017.


	\bibitem{XShi2}
	X. Shi, B. D. Anderson, G. Mao, Z. Yang, J. Chen, and Z. Lin, ``Robust localization using time difference of arrivals,'' \textit{IEEE Signal Process. Lett.}, vol. 23, no. 10, pp. 1320--1324, Oct. 2017.


	\bibitem{SBoyd1}
	S. Boyd and L. Vandenberghe, \textit{Convex Optimization.} Cambridge University Press, 2004.


	\bibitem{DWu}
	D. Wu, J. Zhou, and A. Hu, ``A new approximate algorithm for the Chebyshev center,'' \textit{Automatica}, vol. 49, no. 8, pp. 2483--2488, 2013.



	\bibitem{STomic4}
	S. Tomic and M. Beko, ``A robust NLOS bias mitigation technique for RSS-TOA-based target localization,'' \textit{IEEE Signal Process. Lett.}, vol. 26, no. 1, pp. 64--68, Jan. 2019.



	\bibitem{GWang}
	G. Wang, H. Chen, Y. Li, and N. Ansari, ``NLOS error mitigation for TOA-based localization via convex relaxation,'' \textit{IEEE Trans. Wireless Commun.}, vol. 13, no. 8, pp. 4119--4131, Aug. 2014.


	\bibitem{GWang2}
	G. Wang, A. M. C. So, and Y. Li, ``Robust convex approximation methods for TDOA-based localization under NLOS conditions,'' \textit{IEEE Trans. Signal Process.}, vol. 64, no. 13, pp. 3281--3296, Jul. 2016.
	


	\bibitem{FYin1}
	F. Yin, C. Fritsche, F. Gustafsson, and A. M. Zoubir, ``TOA based robust wireless geolocation and Cramer-Rao lower bound analysis in harsh LOS/NLOS environments,'' \textit{IEEE Trans. Signal Process.}, vol. 61, no. 9, pp. 2243--2255, May 2013.



	\bibitem{ZQLuo}
	Z.-Q. Luo, W.-K. Ma, A. M.-C. So, Y. Ye, and S. Zhang, ``Semidefinite relaxation of quadratic optimization problems,'' \textit{IEEE Signal Process. Mag.}, vol. 27, no. 3, pp. 20--34, May 2010.



	\bibitem{MGrant}
	M. Grant and S. Boyd, ``CVX: MATLAB software for disciplined convex programming, version 2.1.'' Accessed: Sep. 11, 2021. [Online]. Available: \url{http://cvxr.com/cvx}


	\bibitem{WJZeng}
	W.-J. Zeng and H. C. So, ``Outlier-robust matrix completion via $\ell _p$-minimization,'' \textit{IEEE Trans. Signal Process.}, vol. 66, no. 5, pp. 1125--1140, Mar. 2018.

	
	

\end{thebibliography}

\end{document}